\documentclass{elsart}

\usepackage{natbib}

\usepackage{epsfig}

\usepackage{amssymb}

\begin{document}

\begin{frontmatter}



\title{The broad-line radio galaxy story}


\author{Ilse van Bemmel}

\ead{bemmel@astro.rug.nl}

\address{European Southern Observatory, Karl-Schwarzschild-Str.\,2, 
D--85748 Garching}
\address{Kapteyn Astronomical Institute, P.O.Box 800,
NL--9700\,AV Groningen}

\begin{abstract}
% Text of abstract
In this paper I discuss the issue of the so-called 25\,$\mu$m--peakers, 
which were discovered with IRAS, and consist almost solely of
broad-line radio galaxies. I find that this peak is caused by the
absence of colder dust that emits at 60\,$\mu$m and not by an excess
of hot dust, as suggested by the name. On optical images, the
25\,$\mu$m peakers show a lack of extended dust. The peakers are
consistent with being in a later evolutionary stage, in which all the
extended dust has been formed into stars and the dust torus is
smaller. If the 60\,$\mu$m dust is heated by stars, there might be a
correlation between the star-formation rate and the power of the AGN. 

\end{abstract}


\end{frontmatter}

% main text from here

\vspace{-10mm}

\section{A prelude}
\vspace{-5mm}
The first galaxies with nuclear activity were studied in detail
by Seyfert in the 1940's (Seyfert 1943). Now, over 50 years later,
their numbers have increased by several orders of magnitude, revealing
a rich diversity. In their attempt to reach a physical understanding
of active galaxies, people have proposed unified schemes (e.g. Barthel
1989), which combine several apparently different classes into one,
depending on the orientation of a central symmetry 
axis. A dusty torus causes anisotropy at wavelengths were the dust is
optically thick, but is transparent to hard X-rays and radio
emission. This easily explains many differences between Seyfert\,1 and
Seyfert\,2 galaxies, and also between radio galaxies and radio quasars.
However, there are many open questions, e.g. the presence and nature
of type\,2 radio-quiet quasars, the absence of broad emission lines in
some objects (e.g. Hill et al. 1996). In this paper I will study one 
specific problem, that of the 25\,$\mu$m peakers. These are broad-line 
radio galaxies (BLRGs) which have a clear peak in their spectral energy
distribution around 25\,$\mu$m, whereas all other radio-loud AGN have this
peak at longer wavelengths (60--100\,$\mu$m).

\vspace{-5mm}
\section{The problem starts}
\vspace{-6mm}
The properties of the obscuring torus 
have been modeled by Pier \& Krolik (1992) and
Granato \& Danese (1994). Both groups conclude that the dust should
be optically thin in the far-infrared, at wavelengths larger than
$\sim$80\,$\mu$m. Combining this result with the unified models, one
can predict that narrow-line radio galaxies (NLRGs) and radio quasars
(QSRs) of comparable radio power should emit comparable far-infrared
power. No conclusive result was reached when comparing NLRGs and QSRs
in the far-infrared, either with IRAS (Heckman et al. 1994), or with
ISO (van Bemmel et al. 2000). However, one would expect the general
shape of the far-infrared SEDs to be the same for all objects that 
belong to one class, which is true for NLRGs and QSRs.

\begin{figure*}
	\resizebox{14cm}{!}{\includegraphics{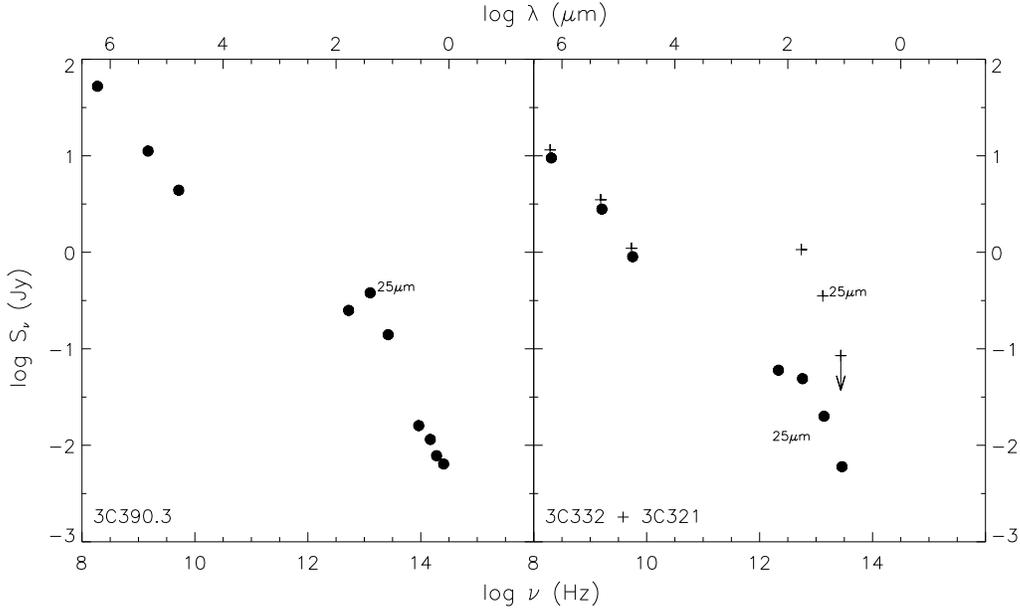}}
	\hfill
	\parbox[b]{14cm}{
	\vspace{-3mm}
	\caption{The  SED in restframe for the BLRG 3C\,390.3 (peaker), and 
	the SEDs of BLRG 3C\,332 (filled circles) and and NLRG 3C\,321
	(plusses), both non-peakers. The 25\,$\mu$m points are marked in the
	plot.}  
	}
\end{figure*}

Broad line radio galaxies were discovered and classified first in the
1960's, when the first radio surveys were done. They are typically
galaxies with a bright nucleus and extended radio structures (like
NLRGs), but a quasar spectrum in the optical. In the light of unified
models, there seems to be an easy way to fit them in; BLRGs can simply
be intermediate angle quasars. The dust torus is seen at an angle that
just obscures the central source (and therefore enables us to see the
host galaxy), but part of the broad-line region is un-obscured. 

As discussed above, one would
expect that the infrared SEDs are the same for NLRGs, BLRGs and
QSRs. However, there is a group of BLRGs which have a peak in
their SED at 25\,$\mu$m, instead of 60--100\,$\mu$m as in NLRGs and
QSRs (see Fig.~1). This seems to be in contradiction with expectations of
the unification schemes, unless these 25\,$\mu$m peakers are somehow
different. So far no NLRGs or QSRs are know with a clear peak in their
SED at 25\,$\mu$m. 
%In Fig.~1 a comparison of the SEDs is shown for 3C\,390.3 
%a 25\,$\mu$m peaker, 3C\,332 a 'normal' BLRG and 3C\,321 a NLRG. 

\begin{figure*}
	\resizebox{7cm}{!}{\includegraphics{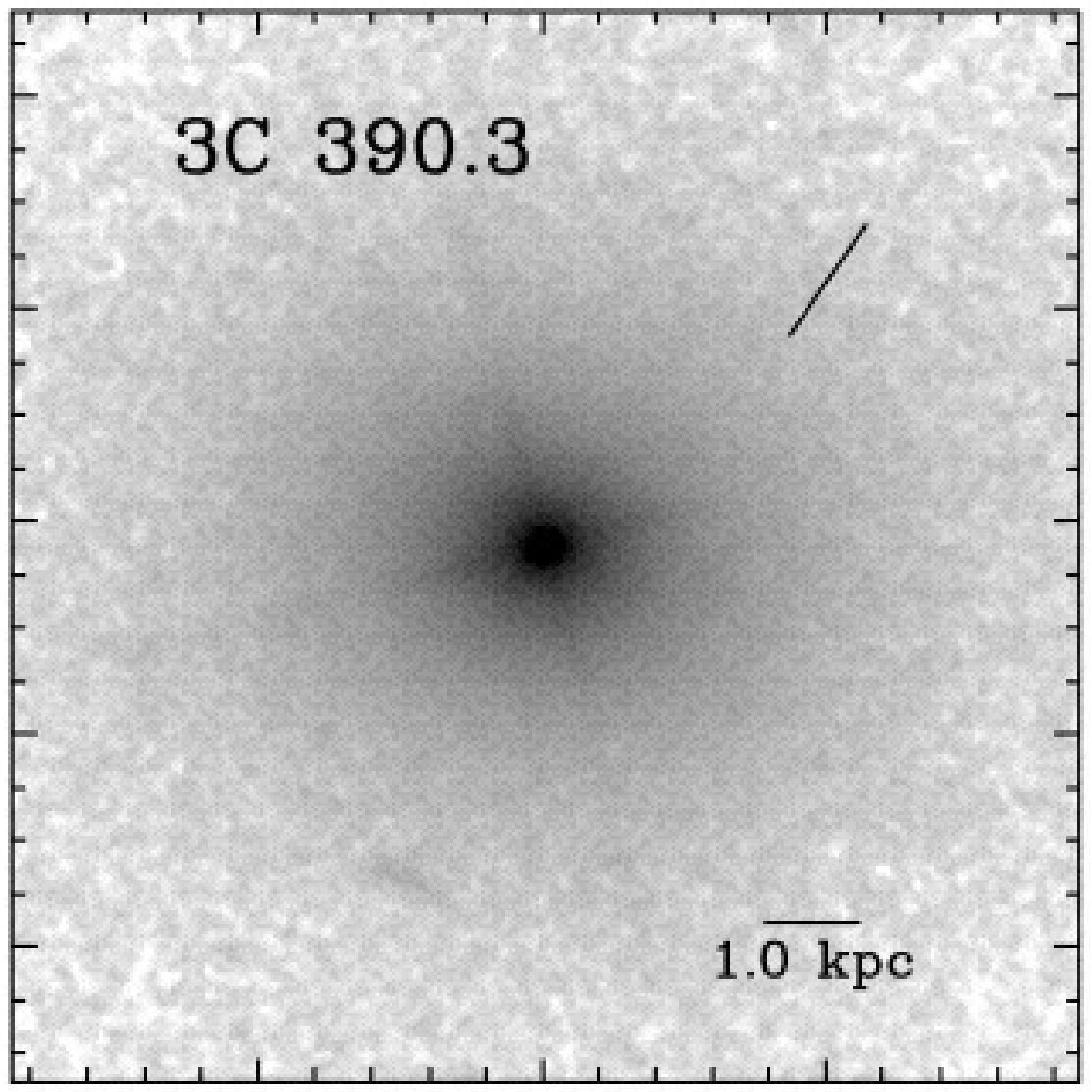}}
	\resizebox{7cm}{!}{\includegraphics{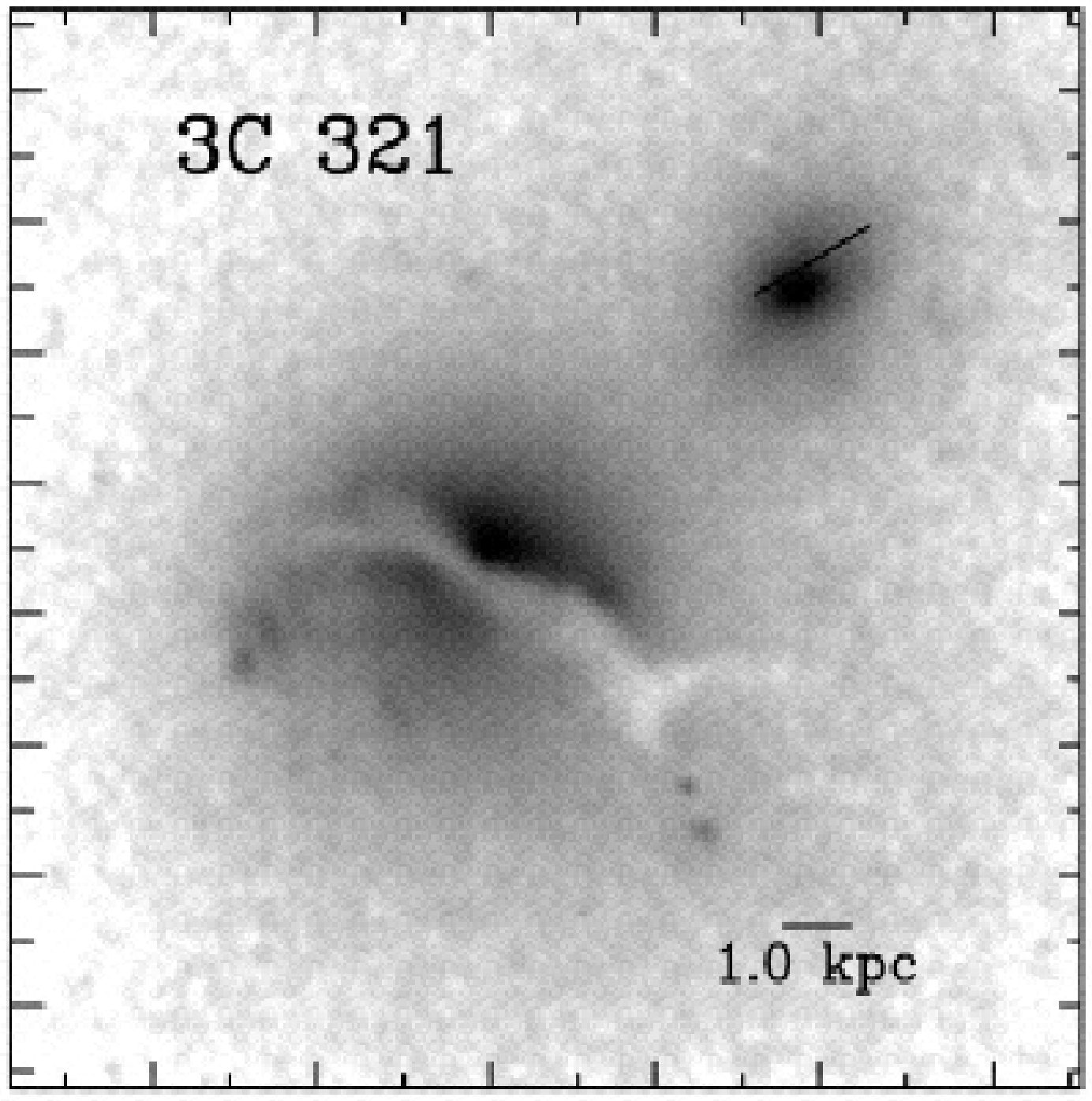}}
	\hfill
	\parbox[b]{14cm}{
	\caption{Optical HST images of 3C\,390.3 and 3C\,321. The
	black lines indicate the radio axis and the image scale
	(de Koff et al. 1996).}
	}	
\end{figure*}

\vspace{-7mm}
\section{Searching for clues}
\vspace{-5mm}
Data were obtained for a limited sample of BLRGs with ISOPHOT on board
ISO (Lemke et al. 1996, Kessler et al. 1996), in the bands 12, 25, 60,
90 and 160\,$\mu$m, using the P1, P2, C1 
and C2 detectors. The observations were made in raster mode, to
avoid the known problems with chopping, which are especially serious
for these faint sources. The reduction of the C1 and C2 data is
described in van Bemmel et al. (2000), and the reduction of the P1 and
P2 data will be described in a future paper (van Bemmel et al. 2001). 
From additional ISOPHOT and literature data, two samples were
selected, one of 25\,$\mu$m peakers and one of normal NLRGs and BLRGs. 
Due to the limited amount of infrared data, the comparison sample
consists largely of NLRGs.

Data in other spectral regimes were extracted from the NASA/IPAC
Extragalactic Database (NED) and the literature. The exact
references for all data will also be given in the future paper (van
Bemmel et al. 2001). The two samples are listed in Table~1, which
gives their classification, infrared powers at 25 and 60\,$\mu$m and
radio power at 178\,MHz. 

\vspace{-5mm}
\section{Lining up the facts}
\vspace{-5mm}

\begin{table}
%\leavevmode
\footnotesize
\begin{center}
\begin{tabular}{|llccc|llccc|}
\hline
\sf
Name&  ID & $P_{25}$&  $P_{60}$& $P_{178}$ & Name& ID & $P_{25}$&  $P_{60}$& $P_{178}$ \\
\hline
\vspace{-2mm}
3C\,33.1& B 	& 24.41	& 24.36	&  --	& 3C\,61.1& N	& 23.97 & 24.59 & 27.39	\\
\vspace{-2mm}
3C\,98	& B	&23.19	&$<$23.15&25.92	& 3C\,79& N	& 24.95	& 25.40	& 27.52	\\
\vspace{-2mm}
3C\,109	& B	& 25.78	& 25.36	& 27.60	& 3C\,111& B	& 24.01	& 24.16	& 26.20	\\
\vspace{-2mm}
3C\,234	& B	& 25.29	& 25.14	& 27.38	& 3C\,321& N	&24.84	& 25.30	& 26.31	\\
\vspace{-2mm}
3C\,382	& B	& 23.78	& 23.82	&  --	& 3C\,327& N	&24.80	& 25.16	& 26.96	\\
\vspace{-2mm}
3C\,390.3&B 	& 24.36	& 24.15	& 26.50	& 3C\,332& B	& 23.99	& 24.36	& 26.63	\\
\vspace{-2mm}
3C\,445	& B	& 24.32	& 24.21	& 26.16	& 3C\,381& B?	& 24.31	& 24.48	& 26.92	\\
\vspace{-2mm}
	&	&	&	&	& 3C\,403& N 	& 24.17	& 24.47	& 26.27	\\
%\vspace{-2mm}
	&	&	&	&	& 3C\,405& N	& 24.70	& 25.25	& 28.75	\\
\hline
{Average}	& &24.45 &$<$24.31& 26.71& {Average}& & 24.42	& 24.80	& 26.99	\\
\hline
\end{tabular}

\vspace{2mm}

\parbox[b]{14cm}{
\caption{Luminosity densities in W\,Hz$^{-1}$ for the peaker sample
(left) and the non-peaker sample (right). Averages for each sample are
given for comparison. The second column gives the nature of
the object, B for BLRG and N for NLRG.} 
}
\end{center}
\end{table}
\normalsize

From here I will refer to the BLRGs with a peak at 25\,$\mu$m as the
peaker sample, and the comparison sample as the non-peaker sample.
For both samples literature data were studied from X-ray, optical,
infrared and radio observations, with 
different observational techniques (imaging, photometry, spectroscopy
and/or polarimetry). There are always objects lacking data, but in
general there were no immediate differences to be found when comparing
the average luminosities over this spectral range and the spectral shape.
However, a closer comparison reveals that there is no evidence at all
for a peak at 25\,$\mu$m, as suggested by the name, it rather seems
that these objects display a lack of 60\,$\mu$m emission. This means 
there is less cold dust in 25\,$\mu$m peakers than in non-peakers.

The 60\,$\mu$m dust can either be heated directly by the AGN (Hes et
al. 1995) or by stars; many star-burst galaxies have a peak in their
SED at 60\,$\mu$m (Calzetti et al. 2000). This leaves two options:
either the peakers have weaker AGN, so the dust is not hot enough to
radiate at 60\,$\mu$m, or there is not enough dust in the host
galaxy to detect it. The first option can be tested by comparing the
178\,MHz power of the two samples. Since the 178\,MHz emission is
isotropic, thus independent on the amount of dust in the host galaxy
and the orientation of the object, the 178\,MHz power should be
significantly weaker for peakers. This is not confirmed in this sample. 

The second option can be tested by comparing optical images. Data from
the HST snapshot survey of the 3CR sample (de Koff et al. 1996) were
used to compare the optical appearance of the two samples (see also
Fig.~2). A blind test was done with three people to classify the
galaxies as non-interacting/dust-free, interacting/dust-rich
or unclear. It turned out that there is a good relation between
the  distortion of the host (including dust lanes) and the shape of
the infrared SED. The peakers are predominantly normal, dust-free
elliptical galaxies, and often reside in sparse environments. The
non-peakers, however, display dust lanes (as seen in Centaurus~A) and
other typical signs of interaction. The dust is usually extended and
often perpendicular to the radio axis, which excludes the AGN as
being the dominant heating source.

\vspace{-5mm}
\section{How to solve the problem}
\vspace{-5mm}

\subsection{Unification}
\vspace{-6mm}
With two groups of BLRGs it might seem that the unification schemes no
longer apply. However, there is an historical pitfall. In the past, an
active galaxy was classified as a radio galaxy when the optical galaxy
was visible. When it was only a point-like source, it would be called
a quasi-stellar object. The names stuck, but with present day
instruments, there are many QSRs with visible host galaxies. Also,
nearby quasars would have been classified as BLRG. A radio study
reveals that there are indeed two groups of BLRGs, one consisting of
low-luminosity quasar counterparts and another of mis-oriented
quasars. Only the last group are BLRGs which fit into unified
models (Dennett-Thorpe et al. 2000). In the sample with 25\,$\mu$m
peakers, there might be more of the first group, while the comparison
sample consists mostly of objects from the second group. This does not
contradict unified models, but it shows that there is more to it than
just orientation.

\vspace{-5mm}
\subsection{Nuclear and extended dust}
\vspace{-6mm}
In the HST images the peakers lack extended dust related to
star-formation, but there is no clear evidence that a central torus is
absent as well. The fact that the broad-line region is visible
indicates that the torus could be smaller in these objects. In the
classical unification schemes there might be confusion between
obscuration of the central parts by a nuclear torus, or by extended
dust in the host galaxy. There is no easy way to tell the difference
between these two geometries, but it implies that NLRG without
extended dust could appear as BLRG, irrespective of the presence of a
nuclear dust torus.

\vspace{-5mm}
\subsection{Evolution}
\vspace{-6mm}
There might be an evolutionary link between 25\,$\mu$m peakers and normal
BLRGs. Dust is the main fuel for both star-formation and the AGN, but
there is a limited supply. Especially when the galaxy is isolated,
there are no ways to re-fuel the host galaxy and star-formation will cease. 
When the fuel for the AGN runs out as well, one expects to see weaker 
emission lines and lower radio power. In addition, without extended dust, 
these galaxies should have little reddening and low polarization. Studies 
of the polarization, colors, stellar populations and X-ray absorption 
columns will provide more clues on the nature of the peakers.    

Hes et al. (1995) show that there is a correlation between 178\,MHz
power and 60\,$\mu$m power, which would imply that the AGN heats the
cold dust. However, this work indicates that in many objects the
60\,$\mu$m dust is related to star-formation processes, as in
star-burst galaxies. Combining both results,
there could be a relation between the star-formation rate in the host
galaxy and the power of the AGN. If one can confirm this relation in
larger samples, it will provide important insights in the formation
and evolution of both the host galaxy and the central black hole.

\vspace{-5mm}
\section{Live long and happy ....}
\vspace{-5mm}
From observations with IRAS and ISOPHOT I find that there is a number
of BLRGs with a peak in their SED at 25$\mu$m. Not all BLRGs show this
behaviour, which deviates from the typical SED for NLRGs and QSRs. The
peaker sample is compared with a sample of NLRGs and BLRGs which have
a peak at 60--100\,$\mu$m. I find no excess of hot dust in the
25\,$\mu$m peakers, but a lack of cold dust, which is the
dust heated by (hot) stars. This might be related to the evolutionary
state of the object, being an older NLRG in which the extended dust
has been depleted by star-formation and the nuclear dust torus is
smaller. The fact that two classes of BLRG seem to exist is not
contradicting the unification schemes, however, it adds an important
new dimension. 

Finally, the name 25\,$\mu$m peaker should be
abandoned as soon as possible, as it is misleading, but so far I have
not found a good alternative.

\end{document}